\documentclass[prd,superscriptaddress,amsfonts,amssymb,amsmath,showpacs]{revtex4-2}
\usepackage{comment}
\usepackage{bm}
\usepackage{amsfonts}
\usepackage{latexsym}
\usepackage[latin1]{inputenc}
\usepackage{graphicx}
\usepackage{amsmath}
\usepackage{palatino}
\usepackage{mathpazo}
\usepackage{textcomp}
\usepackage{float}
\usepackage{booktabs}
\usepackage{dcolumn}
\usepackage{booktabs}
\usepackage{multirow}
\usepackage{hyperref}
\hypersetup{
    colorlinks=true,
    linkcolor=purple,
    filecolor=magenta,      
    citecolor=blue
}
\usepackage{amsmath}
\usepackage{xcolor}
\usepackage{orcidlink}
\usepackage[caption=false]{subfig}
\usepackage{commath}
\makeatletter
\long\def\dddddot#1{%
  {\mathop {#1}\limits ^{\vbox to-1.4\ex@ {\kern -\tw@ \ex@ \hbox {\normalfont .....}\vss }}}%
}
\long\def\multidots#1#2{%
  \count@=0
  {{\mathop {#2}\limits ^{\vbox to-1.4\ex@ {\kern -\tw@ \ex@ \hbox {\normalfont %
  \loop%
  \ifnum#1>\count@%
  .%
  \advance\count@ by1%
  \repeat%
  }\vss }}}}%
}
\makeatother

\begin{document}

\color{black}       

\title{ Accelerating cosmological models in $f(Q)$ gravity and the phase space analysis}

\author{S.A. Narawade\orcidlink{0000-0002-8739-7412}}
\email{shubhamn2616@gmail.com}
\affiliation{Department of Mathematics, Birla Institute of Technology and Science-Pilani, Hyderabad Campus,
Hyderabad-500078, India.}

\author{Shashank P. Singh\orcidlink{0000-0002-7497-7020}}
\email{shashanksingh23101@gmail.com}
\affiliation{Department of Mathematics,
Birla Institute of Technology and Science-Pilani, Hyderabad Campus, Hyderabad-500078, India.}

\author{B. Mishra\orcidlink{0000-0001-5527-3565}}
\email{bivu@hyderabad.bits-pilani.ac.in}
\affiliation{Department of Mathematics, Birla Institute of Technology and Science-Pilani, Hyderabad Campus, Hyderabad-500078, India.}

\begin{abstract}
{\textbf{Abstract}:} The dynamical aspect of accelerating cosmological model has been studied in this paper in the context of modified symmetric teleparallel gravity, the $f(Q)$ gravity. Initially, we have derived the dynamical parameters for two well known forms of $f(Q)$ such as: (i) log-square-root form and (ii) exponential form. The equation of state (EoS) parameter for the dark energy in the $f(Q)$ gravity in both the models emerges into a dynamical quantity. At present model-I shows the quintessence behavior and behave like the $\Lambda$CDM at the late time whereas model-II shows phantom behaviour. Further, the dynamical system analysis has been performed to determine the cosmological behaviour of the models along with its stability behaviour. For both the models the critical points are obtained and analysed the stability at each critical points with phase portraits. The evolutionary behaviour of density parameters for the matter-dominated, radiation-dominated, and dark energy phases are also shown for both the models. 
\end{abstract}

\maketitle

\section{Introduction}
One of the most important implications of the equivalence principle is the expression of gravity through space-time curvature. The most successful theory to explain gravitational interaction is Einstein's general theory of relativity (GR), and the $\Lambda$CDM (Cold Dark Matter) hypothesis is the concordance cosmological model based on GR. The non-renormalizability GR, the cosmological constant problem, the coincidence problem, the Hubble tension, the $\sigma_ 8$ tension, etc. are some of the theoretical and observational issues that this basic gravitational and cosmological framework faces \cite{Addazi2022, Abdalla2022, Martin2012, Freedman2017, Lusso2019, Lin2020, Perivolaropoulos2022}. After astronomical observations of Supernovae over the past few decades \cite{Riess1998, Perlmutter1999}, a number of ideas have been proposed to modify the GR and take into account various formulations of gravity. Depending on the choice of geometrical approaches, one can classify the theories of gravity into three classes: (i) The first one involves use of GR, free torsion, and metric-compatible connections; (ii) The second class, torsion is connected using a metric-compatible, curvature-free approach, such as the teleparallel GR equivalent \cite{Aldrovandi2013}; and (iii) finally the symmetric teleparallel equivalent of GR, requires a curvature and torsion-free connection that is not metrically compatible \cite{Nester1998}. These three equivalent formulations based on the three different connections are commonly known as The Geometrical Trinity of Gravity \cite{Jimenez2019}. In spite of the fact that these three theories are similar at the level of field equations, however, their modifications could not be comparable at the fundamental level \cite{Altschul2015}.\\

The cosmological and astrophysical implications of symmetric teleparallel gravity developed into coincident GR or $f(Q)$ gravity \cite{Jimenez2018}, are being given significance in recent literature on extended theories of gravity. Various works in the literature suggest that the $f(Q)$ theory is one of the promising alternative formulations of gravity to explain cosmological observations \cite{Valentino2021, Yang2021, Valentino2021a, Valentino2021b, Yang2020}. Soudi et al. shows that gravitational wave polarizations have a significant role in restricting the strong field behavior of the gravitational theories \cite{Soudi2019}. Harko et al. have explored an extension of the symmetric teleparallel gravity, by considering a new class of theories where the non-metricity $Q$ is coupled non-minimally to the matter Lagrangian, in the framework of the metric-affine formalism. As in the standard curvature-matter couplings, this non-minimal $Q$-matter coupling entails the non-conservation of the energy-momentum tensor, and consequently the appearance of an extra force \cite{Harko2018}. Testing against a variety of current cosmological observation data, including Type Ia Supernovae, Pantheon data, Hubble data, etc., has been done in order to place observational constraints on the background behaviour of several $f(Q)$ models \cite{Lazkoz2019,  Ayuso2021, Narawade2022}.  For analysing the behaviour of the model of $f(Q)$ gravity in an anisotropic space time, one can see the Refs. \cite{Koussour2022, Koussour2022a}. The motivation behind the $f(Q)$ gravity is to modify the gravitational interactions in such a way that the late time cosmic phenomena can be well explained. The theory allows additional degrees of freedom beyond that of GR, so that it can lead to new gravitational interactions. In the literature, one can find research works on this gravitational theory \cite{Narawade2022a, Maurya2022, Maurya2022a, Maurya2023, Anagnostopoulos2021, Atayde2021, Frusciante2021, Anagnostopoulos2023, Koussour2022b}, but its limitations and predictions are yet to be investigated, in particular its ability to address the cosmic acceleration issue.\\

Before we consider this gravity as a viable alternative to GR, another crucial issue of $f(Q)$ gravity to be addressed, i.e. the stability of its cosmological models. One among several stability analysis is the dynamical stability analysis \cite{Khyllep2023} that studies the behavior of a model under small perturbations. To be specific, it aims at to find the model coming back to the original state or evolve into a different solution. This analysis further provides an accurate prediction on the behaviour of the model pertaining to the physical scenario and thereby resulted in a robust theoretical framework. One can refer some papers on dynamical system analysis in modified theories of gravity \cite{Bonanno2012, Khyllep2021, Agrawal2022, Duchaniya2022, Samaddar2022,Duchaniya2023, Samaddar2023, Pati2023}.\\ 

The form of the function $f(Q)$ defined in this gravity has a major role on the behaviour of the model. The motivation here is to appropriately choose some physically viable form of $f(Q)$ to address the issue of cosmic acceleration and further to check the robustness of the model by performing the dynamical stability analysis. The result may provide some crucial insights into the viability of $f(Q)$ gravity and its possible contribution in understanding the structure and evolution of the Universe. The paper is organised as follows: In Sec. \ref{Sec.II}, the formulation of $f(Q)$ gravity as well its field equations are presented. In Sec. \ref{Sec.III}, two cosmological models are given with some well defined form of $f(Q)$. The dynamical system analysis has been performed in Sec. \ref{Sec.IV} to obtain the critical points and its behaviour. The results and conclusions are noted in Sec. \ref{Sec.V}.

\section{Field Equations of $f(Q)$ Gravity}\label{Sec.II}
We will briefly discuss the $f(Q)$ gravity and its derivation to obtain the field equations. The action of $f(Q)$ gravity \cite{Atayde2021},
\begin{equation}\label{eq:1}
S=\int \frac{1}{2}\left[Q+ f(Q)\right]  \sqrt{-g} d^4 x+\int \mathcal{L}_m \sqrt{-g} d^4 x,   
\end{equation}
where $g$ denotes the determinant of the metric tensor ($g_{\mu \nu}$) and $\mathcal{L}_m$ denotes the matter Lagrangian. The covariant derivative of the metric tensor is the non-metricity tensor,  which can be expressed as,
\begin{equation}\label{eq:2}
Q_{\alpha\mu\nu}=\nabla_{\alpha}g_{\mu\nu}\neq 0~.
\end{equation}
and the two traces are,
\begin{equation}\label{eq:3}
Q_\alpha=Q_{\alpha~~\mu}^{~~\mu} \quad \text { and } \quad \tilde{Q}^{\alpha}=Q_{\mu}^{~~\alpha\mu}.     
\end{equation}
The general affine connection can be decomposed into Levi-Civita connection \cite{Hehl1995, Ortin2015} $\left(\left\{_{~~\mu \nu}^\lambda\right\}\right)$, contortion $\left(K_{~~\mu \nu}^\lambda\right)$ and disformation $\left( L_{~~\mu v}^\lambda\right)$ as, 
\begin{equation}\label{eq:4}
\Gamma_{~~\mu \nu}^\lambda= \left\{_{~~\mu \nu}^\lambda \right\}+K_{~~\mu \nu}^\lambda+L_{~~\mu \nu}^\lambda,
\end{equation}
where
\begin{equation*}
 \left\{_{~~\mu \nu}^\lambda \right\}=\frac{1}{2} g^{\lambda \alpha}\left(\partial_\mu g_{\alpha \nu}+\partial_\nu g_{\alpha \mu}-\partial_\alpha g_{\mu \nu}\right),~~~~~~~~~~~~~K_{~~\mu \nu}^\lambda =\frac{1}{2}\left(T_{~~\mu\nu}^\lambda+T_{\mu~~\nu}^{~~\lambda}+T_{\nu~~\mu}^{~~\lambda}\right), 
 \end{equation*}

 \begin{equation*}
 L_{~~\mu \nu}^\lambda =\frac{1}{2}\left(Q_{~~\mu\nu}^{\lambda}-Q_{\mu~~\nu}^{~~\lambda}-Q_{\nu~~\mu}^{~~\lambda}\right),~~~~~~~~~~~~~T_{~~\mu\nu}^\lambda =\Gamma_{~~\mu\nu}^{\lambda}-\Gamma_{~~\nu\mu}^{\lambda}.
\end{equation*}
The connection attributed to torsion and curvature vanishes on the so-called \textit{Symmetric Teleparallel Equivalent to General Relativity} (STEGR)  \cite{Jimenez2018}. The components of the connection in Eq. \eqref{eq:4} can be rewritten as,
\begin{equation*}
\Gamma_{~~\mu \nu}^\lambda=\frac{\partial y^{\lambda}}{\partial \xi^{\rho }}\partial _{\mu}\partial _{\nu}\xi ^{\rho }.
\end{equation*}
In the above equation, $\xi ^{\lambda}=\xi ^{\lambda}(y^{\mu})$ is an invertible relation and $\frac{\partial y^{\lambda}}{\partial \xi ^{\rho}}$ is the inverse of the corresponding Jacobian \cite{Jimenez2020}. This situation is called a coincident gauge, where there is always a possibility of getting a coordinate system with connections $\Gamma_{~~\mu \nu}^\lambda$ equaling zero. Hence, in this choice, the covariant derivative $\nabla _{\alpha }$ reduces to the partial derivative $\partial _{\alpha }$ i.e. $Q_{\alpha \mu \nu }=\partial _{\alpha }g_{\mu \nu }=\frac{\partial g_{\mu v}}{\partial x^\alpha}$. Thus, it is clear that the Levi-Civita connection $\left\{_{~~\mu \nu}^\lambda \right\}$ can be written in terms of the disformation tensor $L_{\ \mu \nu }^{\alpha }$ as $\left\{_{~~\mu \nu}^\lambda \right\}=-L_{\ \mu \nu }^{\alpha }$.\\
The conjugate of non-metricity called as superpotential is given by,
\begin{equation}\label{eq:5}
P^{\alpha}_{~\mu \nu} \equiv -\frac{1}{4}Q^{\alpha}_{~\mu \nu} + \frac{1}{4}\left(Q^{~\alpha}_{\mu~\nu} + Q^{~\alpha}_{\nu~~\mu}\right) + \frac{1}{4}Q^{\alpha}g_{\mu \nu} - \frac{1}{8}\left(2 \tilde{Q}^{\alpha}g_{\mu \nu} + {\delta^{\alpha}_{\mu}Q_{\nu} + \delta^{\alpha}_{\nu}Q_{\mu}} \right).
\end{equation}
The energy momentum tensor is,
\begin{equation}\label{eq:6}
T_{\mu \nu}=-\frac{2}{\sqrt{-g}} \frac{\delta \sqrt{-g} \mathcal{L}_m}{\delta g^{\mu \nu}},
\end{equation}
Varying action \eqref{eq:1} with respect to the metric tensor, we can get the field equations of $f(Q)$ gravity as \cite{Atayde2021},
\begin{equation}\label{eq:7}
\frac{2}{\sqrt{-g}} \nabla_{\alpha}\left(\sqrt{-g}P_{~\mu \nu}^{\alpha}+\sqrt{-g}f_{Q}P_{~\mu \nu}^\alpha\right)+\frac{g_{\mu\nu}}{2} \Big[Q+f(Q)\Big]+\Big[1+f_{Q}\Big]\left(P_{\mu\alpha\beta}Q_{\nu}^{~~\alpha \beta}-2Q_{\alpha\beta\mu}P^{\alpha\beta}_{~~\nu}\right)=-T_{\mu \nu},
\end{equation}
where $f_Q$ is the derivative of $f(Q)$ with respect to the non-metricity scalar $Q$. Now, to frame the cosmological model, we consider the FLRW space time,
\begin{equation}\label{eq:8}
ds^2=-dt^2+a(t)^2(dx^2+dy^2+dz^2),
\end{equation}
where $a(t)$ represents the scale factor, the non-metricity scalar corresponding to the metric \eqref{eq:8} can be obtained as $Q=6H^{2}$, where $H=\frac{\dot{a}}{a}$ is the Hubble parameter that measures the expansion rate of the Universe. The energy momentum tensor is that of perfect fluid and can be written as,
\begin{equation}\label{eq:9}
T_{\mu\nu}=(\rho+p)u_{\mu}u_{\nu}+pg_{\mu\nu},
\end{equation}
where $p$, $\rho$ respectively be the pressure and energy density. Now, the field equations of $f(Q)$ gravity in perfect fluid can be obtained as \cite{Atayde2021},
\begin{eqnarray} 
\rho &=& \frac{1}{2}(Q+2Qf_{Q}-f)~,\label{eq:10}\\
p &=& \frac{1}{2}(f-Q-2Qf_{Q})-2\dot{H}(2Qf_{QQ}+f_{Q}+1)~.\label{eq:11}
\end{eqnarray}
We denote $f(Q)=f$ and $f_Q=\frac{\partial f}{\partial Q}$ and also we consider that the Universe is filled with dust and radiation fluids. Hence, 
\begin{eqnarray*}
\rho=\rho_{m}+\rho_{r};~~~~~~~~~~~~p=\frac{1}{3}p_{r},
\end{eqnarray*}
with $\rho_{m}$ and $\rho_{r}$ represents the energy density for the matter and radiation phase respectively. Then from Eqns. \eqref{eq:10} and \eqref{eq:11}, we get
\begin{eqnarray}
3H^{2} &=& \rho_{total} = \rho_{r} + \rho_{m} + \rho_{DE}~,\label{eq:12}\\
2\dot{H} + 3H^{2} &=& -p_{total} = -p_{r} - p_{m} - p_{DE}~,\label{eq:13}
\end{eqnarray}
where $p_m$, $p_r$ respectively denotes the pressure of matter and radiation phase; $\rho_{DE}$ and $p_{DE}$ be the energy density and pressure of dark energy (DE) phase, which can be expressed as,
\begin{eqnarray*}
\rho_{DE} &=& \frac{f}{2} - Qf_{Q},\\
p_{DE} &=& 2\dot{H}(2Qf_{QQ} + f_{Q}) + Qf_{Q}-\frac{f}{2}.
\end{eqnarray*}
To note, the above two equations satisfies the conservation equation of the energy momentum tensor, which can be expressed as, $\dot{\rho}+3H(\rho + p) = 0$. Now, the  total EoS parameter and the EoS parameter due to DE can be obtained respectively as,
\begin{eqnarray}
\omega_{total} = \frac{p_{total}}{\rho_{total}} &=& -1+\frac{\Omega_{m} + \frac{4}{3}\Omega_{r}}{2Qf_{QQ}+f_{Q}+1}~,\label{eq:14}\\
\omega_{DE}= \frac{p_{DE}}{\rho_{DE}} &=& -1 + \frac{4\dot{H}(2Qf_{QQ}+f_{Q})}{f-2Qf_{Q}}~.\label{eq:15}
\end{eqnarray}
The density parameter pertaining to pressure-less matter, radiation and DE respectively denoted as, 
\begin{equation}\label{eq:16}
\Omega_{m} = \frac{\rho_{m}}{3H^{2}},~~~~~~~~~~~~~\Omega_{r} = \frac{\rho_{r}}{3H^{2}},~~~~~~~~~~~~~\Omega_{DE} = \frac{\rho_{DE}}{3H^{2}}~.
\end{equation}
The EoS parameter describes the present state of the Universe. A bunch of cosmological observations recently constrained the current value of the EoS parameter to be,   $\omega=-1.29_{-0.12}^{+0.15}$ \cite{Valentino2016},  $\omega=-1.3$ \cite{Vagnozzi2020}, Supernovae Cosmology Project, $\omega = -1.035_{-0.059}^{+0.055}$ \cite{Amanullah2010}; WAMP+CMB, $\omega = -1.079_{-0.089}^{+0.090}$ \cite{Hinshaw2013}; Plank 2018, $\omega = -1.03\pm 0.03$ \cite{Aghanim2020}  or $\omega=-1.33_{-0.42}^{+0.31}$ \cite{Valentino2021c}.

\section{The $f(Q)$ gravity Models}\label{Sec.III}
To solve the above system and to analyse the behaviour of the dynamical parameters, we consider some form of Hubble parameter $H$ and the function $f(Q)$. The Hubble parameter, $H=\xi+\frac{\eta}{t}$ corresponds to the scale factor, $a(t)=e^{\xi t}t^{\eta}$, known as the hybrid scale factor \cite{Tripathy2015, Mishra2018, Mishra2019, Tripathy2021}. Subsequently, the non-metricity scalar becomes, $Q=6H^2=6\left(\xi+\frac{\eta}{t}\right)^2$. This form of Hubble parameter further provides a time varying deceleration parameter $\left(q=-1+\frac{\eta}{(\xi t+\eta)^2}\right)$, which can simulate a cosmic transition from early deceleration to late time acceleration. The deceleration parameter, $q\approx -1+\frac{1}{\eta}$ when $t\rightarrow0$ whereas $q\approx-1$ when $t\rightarrow \infty$. To realise the positive deceleration at early Universe for the  transient Universe, the scale factor parameter $\eta$ should be $0<\eta<1$. The transition can occur at $t=-\frac{\eta}{\xi}\pm\frac{\sqrt{\eta}}{\xi}$. We will restrict to the positivity of the second term and ignore the negativity. This is because the negativity of the second term would provide negative time, which may lead to unphysical situation at the bigbang scenario. In that case also the parameter $\eta$ restricted to $0<\eta<1$, since the cosmic transit may have occurred at, $t=\frac{-\eta+\sqrt{\eta}}{\xi}$. The jerk parameter, $j=1-\frac{3\eta}{(\eta+\xi t)^2}+\frac{2\eta}{(\eta+\xi t)^3}$. We have used the parameter values for model analysis as $\xi = 0.965$ and $\eta = 0.60$ \cite{Tripathy2015}.\\

\begin{figure}[H]
    \centering
\includegraphics[scale=0.5]{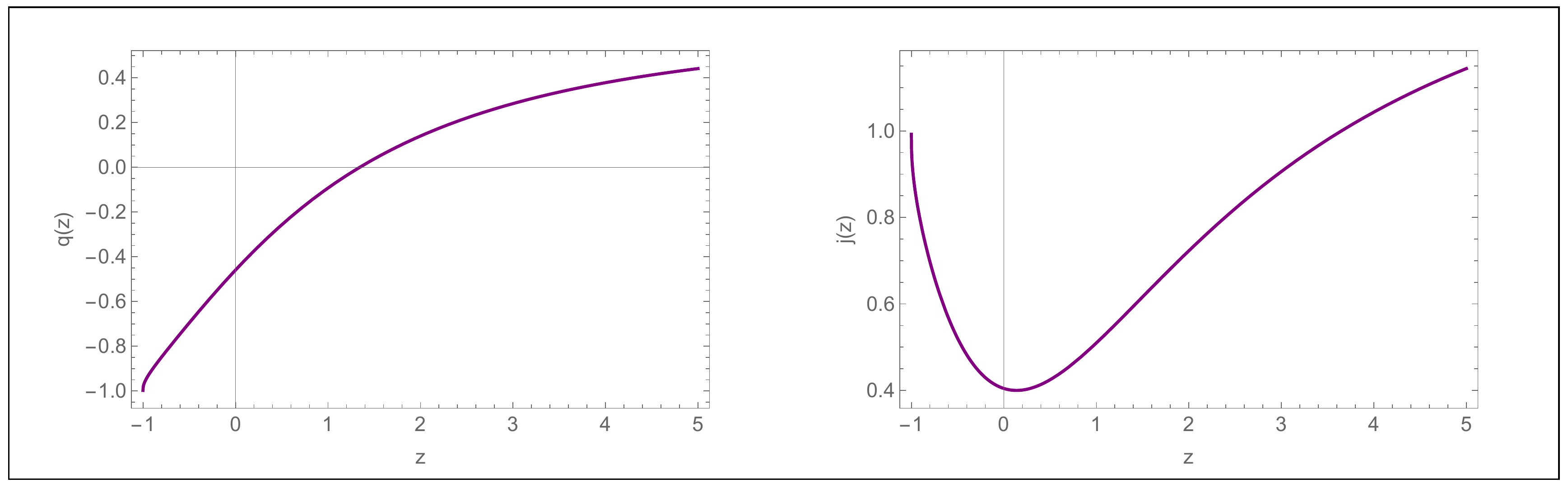}
\caption{Evolution of deceleration parameter (Left Panel) and jerk parameter (Right Panel) in redshift. The parameter scheme: $\xi = 0.965$, $\eta = 0.60$.}
 \label{fig:gp}
\end{figure}
The deceleration parameter shows a transient behaviour with the transition occurs at $z=1.34$. It reduces from early time to late times and at present its value is, $q_0 = -0.46$ [FIG.- \ref{fig:gp} (Left panel)]. The evolution of jerk parameter remains entirely in the range $0<j<1$. It reduces from early time and attained the minimum value at present and eventually converges to $1$ at the late times [FIG.- \ref{fig:gp} (Right panel)].

\subsection{Model-I}
We first consider the logarithmic form of $f(Q)$ \cite{Anagnostopoulos2023} as, 
\begin{equation}\label{eq:17}
f(Q)=nQ_{0}\sqrt{\frac{Q}{\lambda_{1}Q_{0}}}\ln{\frac{\lambda_{1}Q_{0}}{Q}}~,
\end{equation}
where $n$ and $\lambda_{1}>0$ are free parameters; $Q_{0}=6H_{0}^{2}$, where $H_{0}=70.7~kms^{-1}Mpc^{-1}$ \cite{Hotokezaka2019} represents the present value of $H$. For $n=0$, one can recover the GR equivalent model.

\begin{figure}[H]
\centering
\includegraphics[scale=0.5]{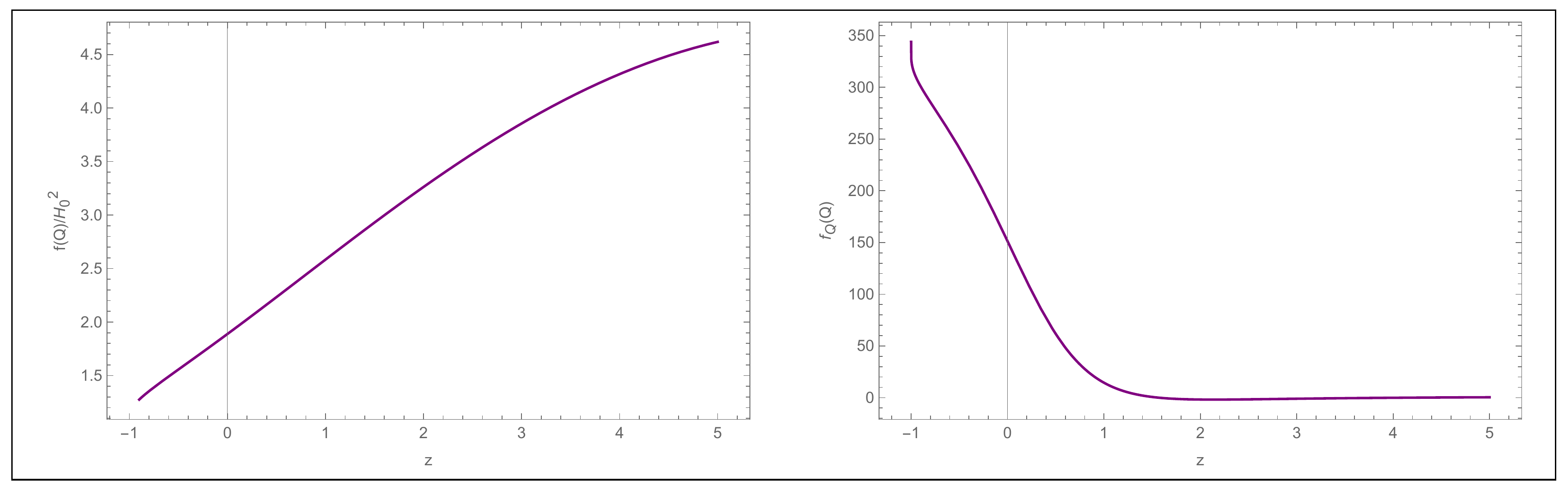}
\caption{Evolution of $\frac{f(Q)}{H_{0}^2}$ (Left Panel) and $f_Q(Q)$ (Right Panel) in reshift for {\bf Model-I}. The parameter scheme: $\xi = 0.965$, $\eta = 0.60$, $n = 1$, $Q_{0} = 29990$ and $\lambda_{1}=0.35$ .}
 \label{fig:I}
\end{figure}

At the outset, we have studied the evolutionary behaviour of the function $f(Q)$ by plotting the graphs $\frac{f(Q)}{H_{0}^2}$ vs $z$ and $f_Q$ vs $z$ where $z$ is redshift \cite{Frusciante2021}. We wish to analyse $\frac{f(Q)}{H_{0}^2}$ and $f_Q$ as functions of redshift since the Hubble parameter is related to the scale factor as $H=\frac{\dot{a}}{a}$. One can see from FIG.- \ref{fig:I} as time passes, $\frac{f(Q)}{H_{0}^2}$ shows a decreasing behavior and gradually vanishes. Whereas, $f_Q$ starts with a lower positive value, increases over time, and remains positive throughout.
Using Eqn. \eqref{eq:17}, Eqns. \eqref{eq:10}, \eqref{eq:11} and \eqref{eq:14} reduces to,
\begin{eqnarray}
p_{total} &=&\frac{\lambda_{1}\Omega_{r}(z+1)^{4}\sqrt{\frac{H^{2}(z)}{\lambda_{1}Q_{0}}}-3\sqrt{6}H^{2}(z)n+\sqrt{6}H(z)n(z+1)H_{z}(z)}{3\lambda_{1}\sqrt{\frac{H^{2}(z)}{\lambda_{1}Q_{0}}}}~,\label{eq:18}\\
\rho_{total} &=& H(z)Q_{0}n\sqrt{\frac{6}{\lambda_{1}Q_{0}}}+\Omega_{r}(z+1)^{4}+\Omega_{m}(z+1)^{3}~, \label{eq:19}\\
\omega_{total} &=& \frac{\lambda_{1}\Omega_{r}(z+1)^{4}\sqrt{\frac{H^2(z)}{\lambda_{1}Q_{0}}}-3\sqrt{6}H^2(z)n+\sqrt{6}H(z)n(z+1)H_{z}(z)}{3\sqrt{\frac{H^2(z)}{\lambda_{1}Q_{0}}}\left(\sqrt{6}\lambda_{1}Q_{0}n\sqrt{\frac{H^2(z)}{\lambda_{1}Q_{0}}}+\lambda_{1}(z+1)^{3} (\Omega_{m}+\Omega_{r}z+\Omega_{r})\right)}~,\label{eq:20}
\end{eqnarray}
\begin{figure}[H]
    \centering
\includegraphics[scale=0.55]{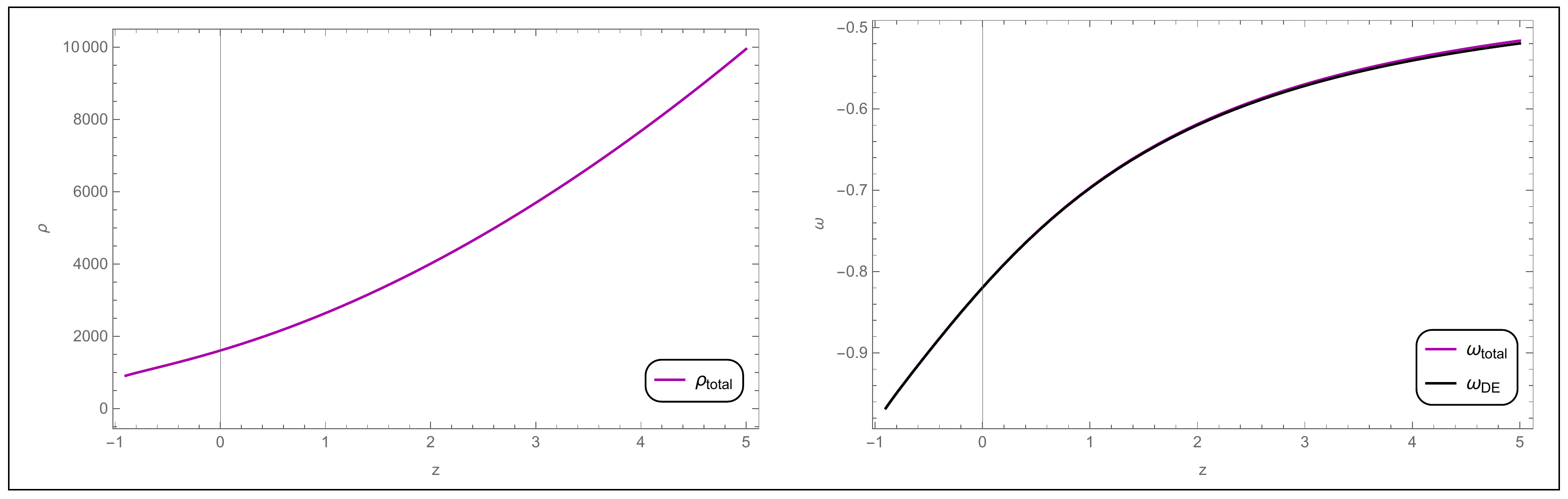}
\caption{Evolution of total energy density (Left Panel) and EoS parameters (Right Panel) in redshift for {\bf Model-I}. The parameter scheme: $\xi = 0.965$, $\eta = 0.60$, $n = 1$ $Q_{0} = 29990$, $\lambda_{1}=0.35$, $\Omega_{m}=0.3$ and $\Omega_{r}=0.00001$.}
\label{fig:II}
\end{figure}
where $H_{z}$ denotes the $\frac{dH(z)}{dz}$. The parameters $\xi$, $\eta$, $n$, $Q_{0}$ and $\lambda_{1}$ determine the evolutionary behavior of total energy density and EoS parameters. The model parameter has been chosen in such a way that a positive energy density can be obtained FIG.- \ref{fig:II} (Left Panel). The total energy density is showing decreasing behavior from early epoch to late epoch and remains positive throughout. The total EoS parameter shows quintessence behavior at present epoch, whereas it converges to $\Lambda CDM$ at late epoch [FIG.- \ref{fig:II} (Right Panel)]. At $z=0$, the value of total EoS parameter observed to be $\approx -0.82$. The behaviour of DE EoS parameter remains almost same with that of total EoS parameter.

\subsection{Model-II}
As a second model, we consider an exponential function of $f(Q)$ \cite{Anagnostopoulos2021} as,
\begin{equation}\label{eq:21}
f(Q)=Qe^{\frac{\mu\lambda_{2}}{Q}}-Q~,
\end{equation}
where $\lambda_{2}$ is the free parameter. In the cosmological framework, the model gives rise to a scenario without $\Lambda CDM$ as a limit, having the same number of free parameters as $\Lambda CDM$. In this model, in certain time period of cosmic history, the term $\frac{\mu}{Q}$ decreases, which ends up making the model as polynomial one. 
\begin{figure}[H]
\centering
\includegraphics[scale=0.5]{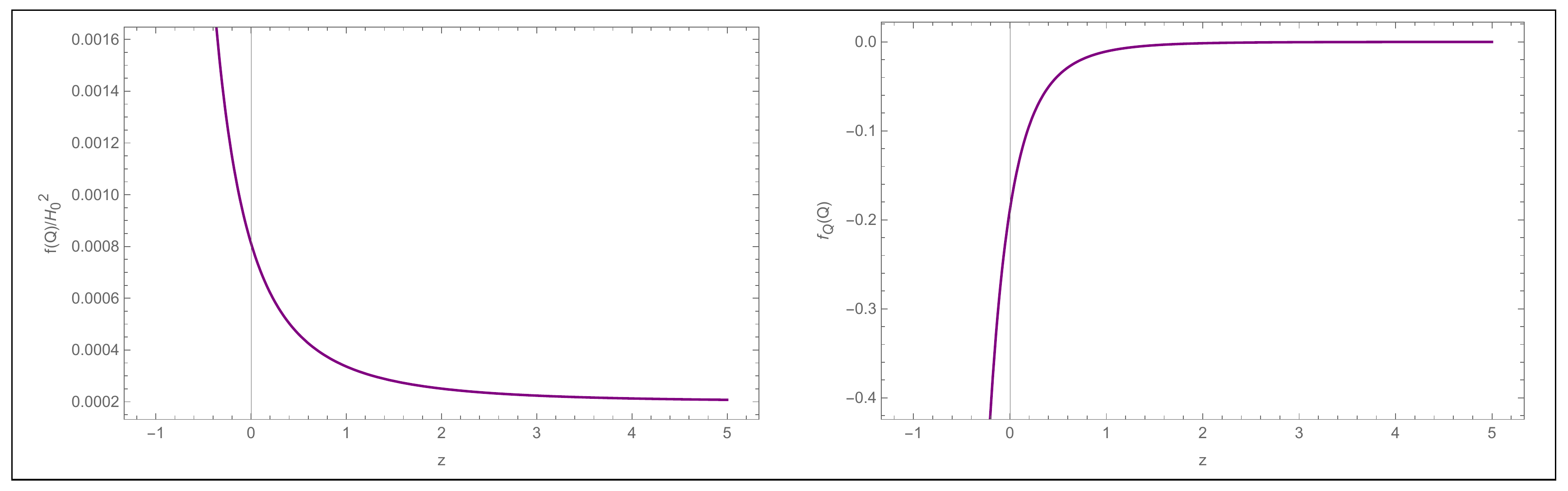}
\caption{Evolution of $\frac{f(Q)}{H_{0}^2}$ (Left Panel) and $f_Q(Q)$ (Right Panel) in redshift for {\bf Model-II}. The parameter scheme: $\xi = 0.965$, $\eta = 0.60$, $\mu = 6.5$ and $\lambda_{2} = 6.5$}
\label{fig:III}
\end{figure}
We can see the behaviour of $\frac{f(Q)}{H_{0}^2}$ and $f_Q$ in FIG.- \ref{fig:III}. In this model also, the normalized non-metricity function $\frac{f(Q)}{H_{0}^2}$ increases with cosmic time due to exponential form of model. However, the functional $f_Q$ becomes a decreasing function of the redshift. The motivation behind these plots is that the non-metricity scalar $Q$ at the FLRW background assumes $6H^{2}$. To analyse the dynamic behavior of the model, the dynamical parameters can be derived from Eqns. \eqref{eq:10}, \eqref{eq:11} and \eqref{eq:14} by incorporating Eqn. \eqref{eq:21} as, 
\begin{eqnarray}\label{eq:22}
p_{total} &= -3H^2(z)+2H(z)(z+1)H_{z}(z)-\frac{e^{\frac{\lambda_{2}\mu}{6H^2(z)}} \left(9H^3(z)\left(3H^2(z)+\lambda_{2}\mu\right)+\lambda_{2}\mu(z+1)H_{z}(z)\left(9H^2(z)+\lambda_{2}\mu\right)\right)}{54H^5(z)}+\frac{1}{3}\Omega_{r}(z+1)^4~,
\end{eqnarray}
\begin{eqnarray}
\rho_{total} &=& \frac{1}{6}e^{\frac{\lambda_{2}\mu}{6H^2(z)}} \left(\frac{\lambda_{2}\mu}{H^2(z)}+3\right)+3H^2(z)+(z+1)^{3}(\Omega_{m}+\Omega_{r}z+\Omega_{r})~,\label{eq:23}\\
\omega_{total} &=& \frac{-3H^2(z)+2H(z)(z+1)H_{z}(z)-\frac{e^{\frac{\lambda_{2}\mu}{6H^2(z)}} \left(9H^3(z)\left(3 H^2(z)+\lambda_{2}\mu\right)+\lambda_{2}\mu(z+1)H_{z}(z)\left(9 H^2(z)+\lambda_{2}\mu\right)\right)}{54H^5(z)}+\frac{1}{3}\Omega_{r}(z+1)^4}{\frac{1}{6} e^{\frac{\lambda_{2}\mu}{6H^2(z)}} \left(\frac{\lambda_{2}\mu}{H^2(z)}+3\right)+3H^2(z)+(z+1)^3(\Omega_{m} +\Omega_{r}z+\Omega_{r})}\label{eq:24}
\end{eqnarray}
\begin{figure}[H]
\centering
\includegraphics[scale=0.55]{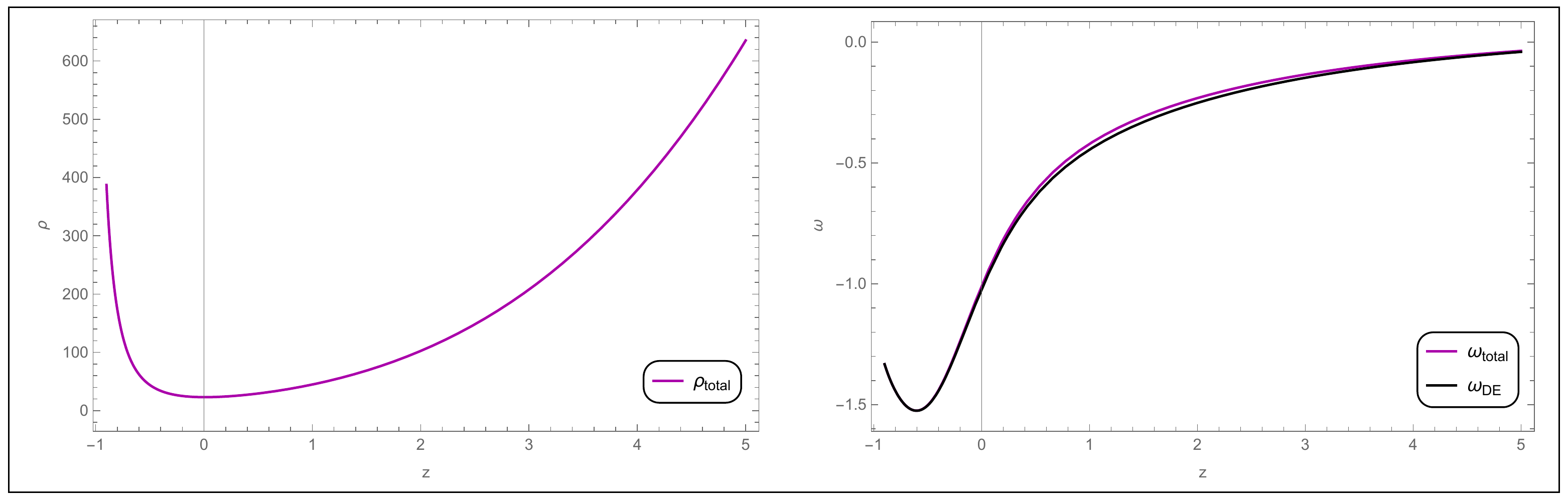}
\caption{Evolution of total energy density (Left Panel) and EoS parameters (Right Panel) in redshift for {\bf Model-II}. The parameter scheme: $\xi = 0.965$, $\eta = 0.60$, $\mu = 6.5$, $\lambda_{2} = 6.5$, $\Omega_{m}=0.3$ and $\Omega_{r}=0.00001$.}
\label{fig:IV}
\end{figure}
FIG.- \ref{fig:IV} (Left Panel) shows the evolutionary behaviour of total energy density that remains positive throughout the evolution and at present time it attains the minimum value. The total EoS parameter remains entirely in the negative domain and decreases from early time. The present value has been obtained as, $\omega_{total}=-1.01$ [FIG.- \ref{fig:IV} (Right Panel)]. At present the DE EoS parameter remains alike to that of total EoS parameter though there was a marginal change noticed at early time.

\section{The Phase Space Analysis}\label{Sec.IV}
\begin{figure}[H]
    \centering
    \includegraphics[scale=0.5]{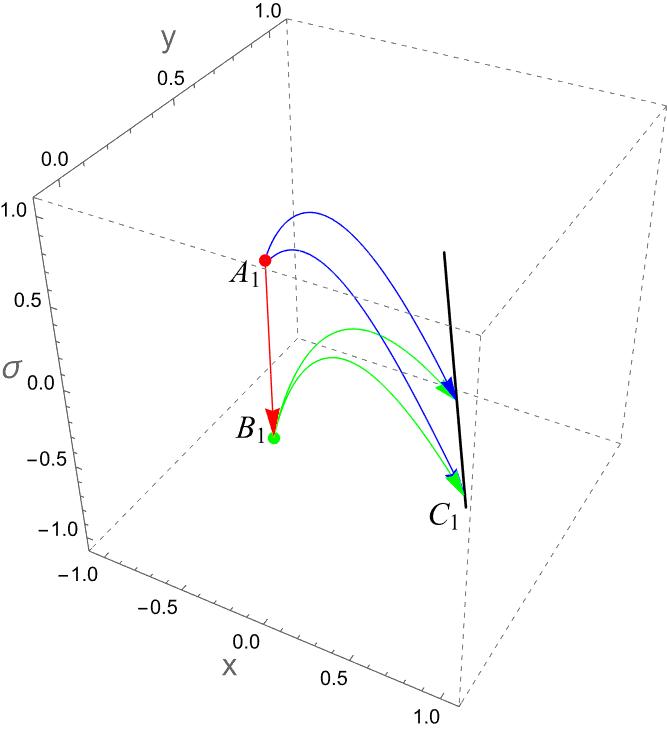}
    \includegraphics[scale=0.7]{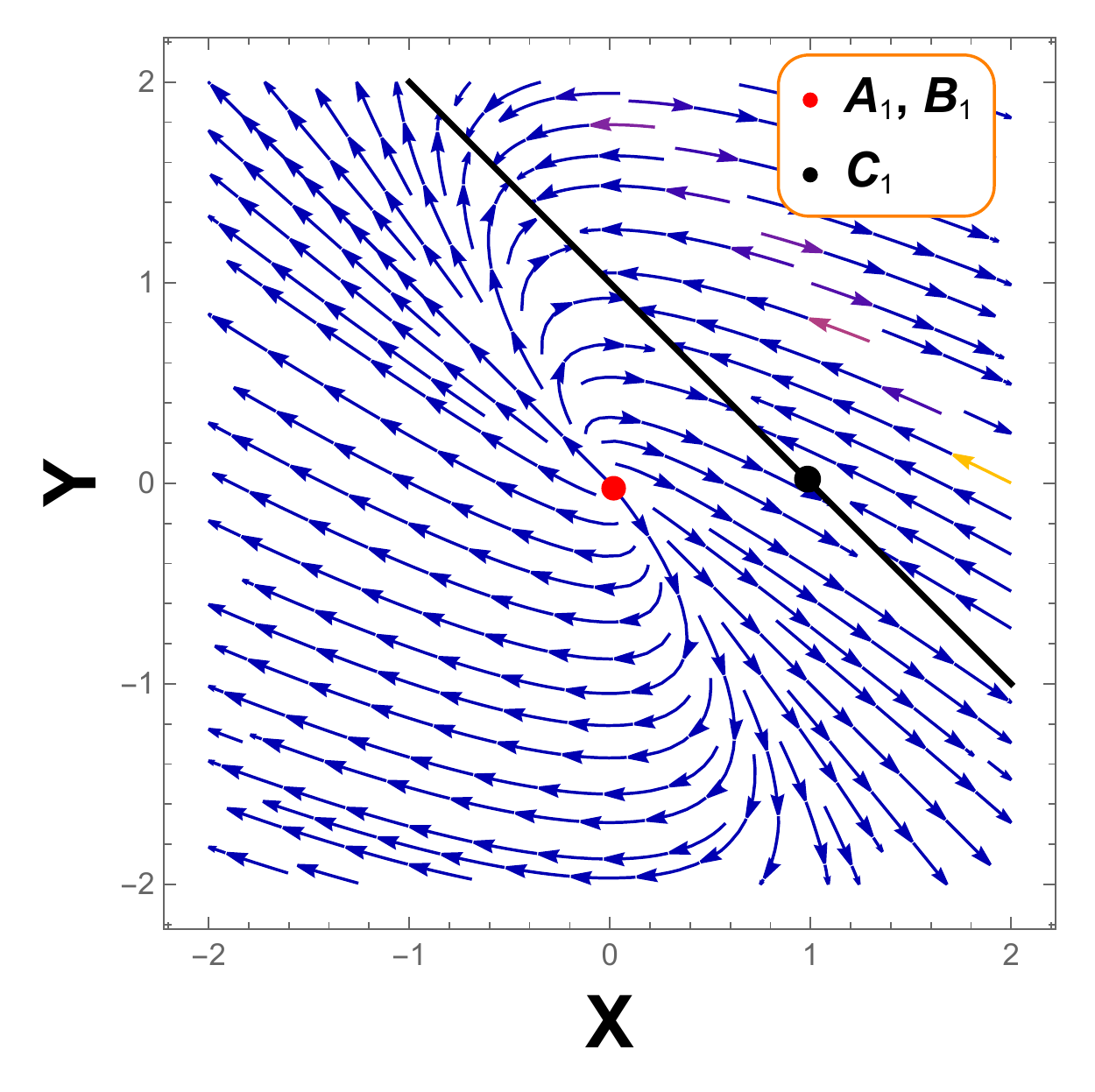}
    \caption{Phase-space trajectories on the $x$-$y$-$\sigma$ plane (Left Panel) and the portrait on the $x$-$y$ plane (Right Panel) for {\bf Model-I}.}
    \label{fig:V}
\end{figure}

\begin{figure}[H]
\centering
\includegraphics[scale=0.7]{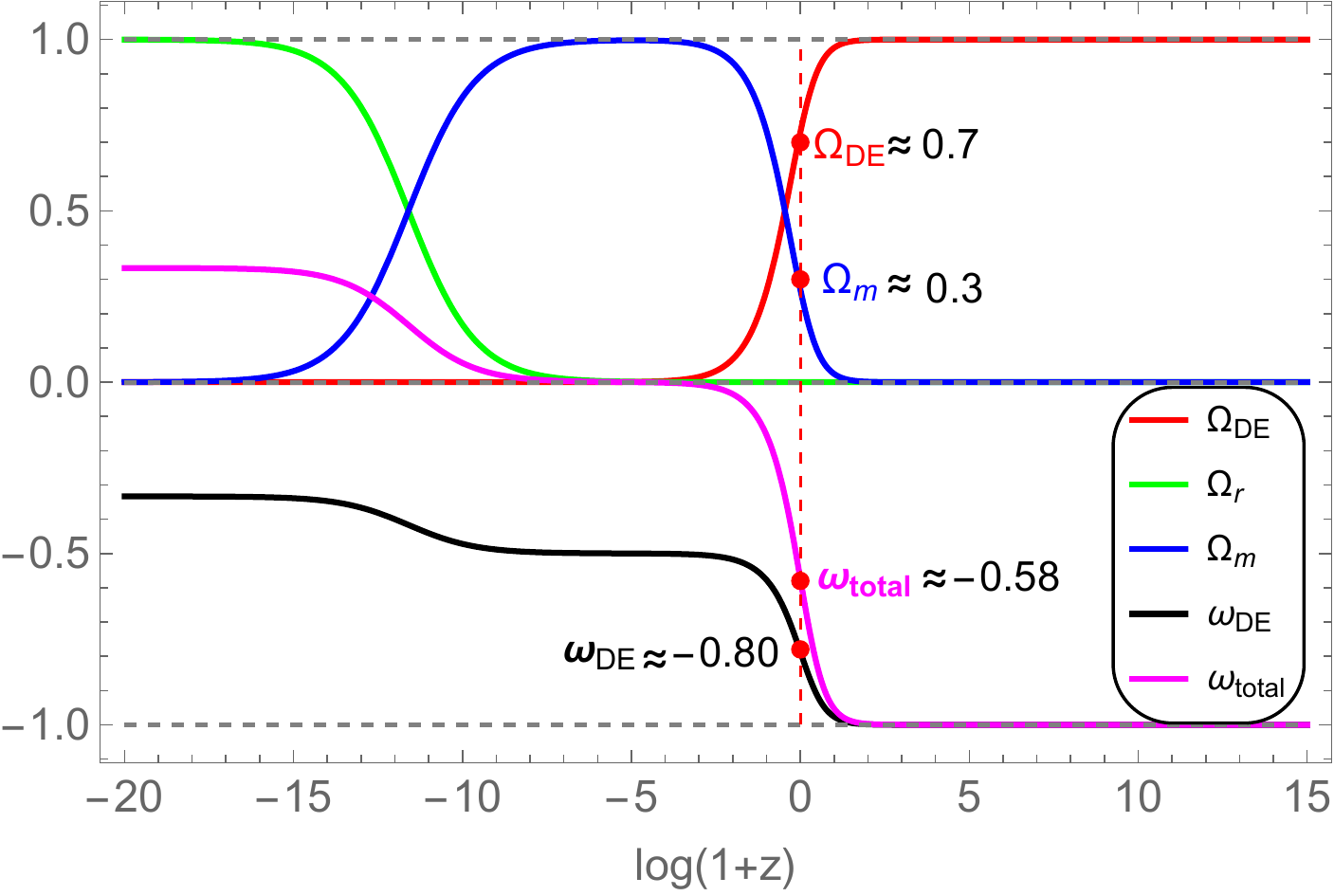}
\caption{Evolution of parameters for {\bf Model-I}. The initial conditions: $x=10^{-15}$, $y=10^{-6}$ and $\sigma=10^{-1}$. The vertical dashed red line denotes the present time.}\label{fig:VI}
\end{figure}

Any group of elements that evolves over time is a dynamical system, whether they are real or even artificial. The dynamical system is based on differential equations associated with time derivatives. So, it is unlike that there exists any universal theory of dynamical systems. The evolution rule that governs the dynamical system should therefore be analyzed in various ways to find its characteristics \cite{Abraham1992, Katok1995, Strogatz2000}. In order to probe the evolutionary dynamics of the theory, we use dynamical systems instead of solving the non-linear differential equations that describes the majority of cosmological models. The stability can be analysed in various methods, some of them are Jacobi stability, Kosambi-Cartan-Chern (KCC) theory or Lyapunov methods. We shall use the Jacobi stability analysis in this problem. We shall perform the dynamical system analysis of the background equations of the two models and will focus on its stability \cite{Bahamonde2018, D'Agostino2018}. To do so, we consider 3-dimensionless parameters, $x$, $y$ and $\sigma$, which may give detailed idea about DE and transform the field equations in terms of the dynamical variables as,

\begin{eqnarray*}
x=\frac{f}{6H^2},\hspace{1cm} y=-2f_{Q},\hspace{1cm} \sigma=\frac{\rho_r}{3H^2}.
\end{eqnarray*}
Subsequently, Eqn. \eqref{eq:16} reduces to,
\begin{eqnarray*}
\Omega_r &=& \sigma,\\
\Omega_{DE} &=& x+y,\\
\Omega_m &=& 1-x-y-\sigma,\\
\frac{\Dot{H}}{H^2} &=& \frac{-(3-3x-3y+\sigma)}{2(2Qf_{QQ}+f_Q+1)}~.
\end{eqnarray*}
Here, prime (~'~) represents differentiation with respect to the number of e-folds of the Universe, $N=ln a$. Then we can differentiate $x$, $y$, and $\sigma$ with respect to $N$ to obtain,
\begin{eqnarray}
x^{\prime} &=& -\frac{\Dot{H}}{H^2}(2x+y) = \frac{(3-3x-3y+\sigma)(2x+y)}{2(2Qf_{QQ}+f_Q+1)},\label{eq:25}\\
y^{\prime} &=&- \frac{\dot{H}}{H^2}(4Qf_{QQ}) = (3-3x-3 y+\sigma)\left[1+\frac{(y-2)}{2(2 Qf_{QQ}+f_Q+1)}\right], \label{eq:26}\\
\sigma^{\prime} &=& -\sigma\left[4+2\frac{\Dot{H}}{H^2}\right]  = \sigma\left[\frac{(3-3x-3y+\sigma)}{(2Qf_{QQ}+f_Q+1)}-4\right]. \label{eq:27}
\end{eqnarray}
Now, we can redefine $\omega_{total}$ and $\omega_{DE}$ as
\begin{eqnarray*}
\omega_{total} &=& -1-\frac{2}{3}\frac{\Dot{H}}{H^2},\\
\omega_{DE} &=& -1-\frac{1}{3(x+y)}\left[y^{'}+\frac{\Dot{H}}{H^2}y\right].
\end{eqnarray*}
For $f(Q)=nQ_{0}\sqrt{\frac{Q}{\lambda_{1}Q_{0}}}~ln\left(\frac{\lambda_{1}Q_{0}}{Q}\right)$ [\textbf{Model-I}], one can express Eqns. \eqref{eq:25}-\eqref{eq:27} in terms of dynamical variables as,
\begin{eqnarray}
 x^{\prime}& =& \frac{(2x+y)(3-3x-3y+\sigma)}{(2-x-y)},\label{eq:28}\\
 y^{\prime} &=& \frac{x(3x+3y-3-\sigma)}{(2-x-y)}, \label{eq:29}\\
 \sigma^{\prime} &=& \frac{2\sigma(\sigma-x-y-1)}{(2-x-y)}.\label{eq:30}
\end{eqnarray}
Moreover, the total EoS parameter and EoS parameter for DE can be written in dynamical variables as,
\begin{eqnarray*}
\omega_{total} &=& -1-\frac{2}{3}\left( \frac{3x+3y-\sigma-3}{2-x-y}\right),\\
\omega_{DE} &=& -\frac{3-\sigma}{3(2-x-y)}.
\end{eqnarray*}
\begin{table}[H]
\renewcommand\arraystretch{1.5}
    \centering
    \begin{tabular}{|c|c|c|c|c|c|c|c|c|c|}
        \hline 
 ~~Name~~ & ~~Point/Curve~~ & ~~$\Omega_{m}~~$ & ~~$\Omega_{r}$~~ & ~~$\Omega_{DE}$~~ & ~~$q$~~ & ~~$\omega_{total}$~~ & ~~$\omega_{DE}$~~& ~~Phase of Universe~~ & ~~Stability~~ \\
\hline \hline
$A_{1}$ & (0, 0, 1) & 0 & 1 & 0 & 1 & $\frac{1}{3}$ & $-\frac{1}{3}$ & Radiation dominated & Unstable Node\\
\hline
$B_{1}$ & (0, 0, 0) & 1 & 0 & 0 & $\frac{1}{2}$ & 0 & $-\frac{1}{2}$ & Matter dominated &  ~~Unstable Saddle~~\\
\hline
$C_{1}$ & ($x$, $1-x$, 0) & 0 & 0 & 1 & -1 & -1 & -1 & ~~DE dominated~~ & Stable Node\\
\hline
\end{tabular}
\caption{Critical Points and the corresponding cosmology for {\bf Model-I}}
\label{table:I}
\end{table}
TABLE \ref{table:I} provides the critical points and the cosmological behaviour at these points. The details description of each critical point has been narrated below. In FIG.- \ref{fig:V}, the $2D$ and $3D$ phase portrait have been given to understand the stability of these points.
\begin{itemize}
 \item \textbf{Critical point $A_1$ (0, 0, 1) :} The
    corresponding EoS parameter and deceleration parameter is $\omega_{total}=\frac{1}{3}$ and $q=1$ respectively. This behaviour of the critical point leads to the decelerating phase
    of the Universe. The density parameters are, $\Omega_{DE}=0$, $\Omega_{m}=0$ and $\Omega_{r}=1$. This critical point is an unstable node because it contains all positive eigenvalues of the Jacobian matrix,
    \begin{eqnarray*}
    \{2,~2,~1\}.
    \end{eqnarray*}
    
    \item \textbf{Critical Point $B_1$ (0, 0, 0) :} The critical point leads to the decelerating phase of the Universe, since the EoS parameter corresponding to this critical point is $\omega_{total}=0$ and deceleration parameter is $q=\frac{1}{2}$. The corresponding density parameter are $\Omega_{DE}=1$, $\Omega_{m}=0$ and $\Omega_{r}=0$. The eigenvalues for corresponding critical point shows positive and negative signature as shown below. The critical point shows unstable saddle behaviour. 
    \begin{equation*}
    \left\{\frac{3}{2},~~\frac{3}{2},~-1\right\}. 
    \end{equation*}

    \item \textbf{Curve of Critical Points $C_1$ (x, 1-x, 0) :} At this point, $\Omega_{DE}=1$, $\Omega_{m}=0$ and $\Omega_{r}=0$, i.e. the Universe shows DE dominated phase. The accelerated DE dominated Universe is confirmed by the corresponding value of the EoS parameter ($\omega_{total}=-1$) and value of the deceleration parameter ($q=-1$). Jacobian matrices with critical points have negative real parts and zero eigenvalues. Further, there is only one vanishing eigenvalue and therefore the dimension of the set of eigenvalues equals its number. As a result, the set of eigenvalues is normally hyperbolic, the critical point associated with it cannot be a global attractor \cite{Aulbach1984, Coley1999}. This critical point, shows stable node behaviour. The corresponding eigenvalues are given below:
    \begin{equation*}
    \{-4,~-3,~~0\}.
    \end{equation*}
\end{itemize}
From the phase portrait [FIG.- \ref{fig:V}], we  can see that the critical point $A_{1}$ is unstable node, where as curve [($x$, $1-x$, 0)] is stable node. The $B_{1}$ is unstable saddle point. FIG.- \ref{fig:V} (Left Panel) describes the trajectories for critical points, where $A_{1}$ is the repeller, so it repels every trajectory and $C_{1}$ is the attractor, so it absorbs every trajectory coming towards it. The $B_{1}$ is saddle therefore, it absorbs the trajectories coming from $A_{1}$ and repel the trajectories originated from itself. FIG.- \ref{fig:V} (Right Panel), one can observe that the stability is not only specific to the single point but also in the entire curve ($x$, $1-x$, $0$). This kind of stability behaviour may be due to the fact that irrespective of the value of the dynamical variable $x$, it exhibits the stable behaviour. It may be due to the nature of the dynamical variable $x$.  The $3D$ portrait shows the trajectory behaviour of the model starting from repeller point $A_{1}$ to the saddle point $B_{1}$ and then it is moving from $B_{1}$ to the stable curve $C_{1}$. Further, the evolution plot for {\bf Model-I} has been given in FIG.- \ref{fig:VI}. From the evolution curve, the present value of DE EoS parameter is $-0.80$ whereas the value obtained using the hybrid scale factor is $-0.82$. Hence in both the approaches, we can get similar value at present time and the Universe shows quintessence behaviour. At present the value of density parameters for DE and matter obtained as $\approx 0.7$ and $\approx 0.3$ respectively.

For, $f(Q)=Qe^{\frac{\mu\lambda_{2}}{Q}}-Q$ [\textbf{Model-II}], we can get the system of differential equations as,
\begin{eqnarray}
x^{\prime} &=& \frac{(3-3x-3y+\sigma)(2x+y)}{2-y+4(x+1)[\ln(x+1)]^2}, \label{eq:31}\\
y^{\prime} &=& \frac{4(x+1)(3-3x-3y+\sigma)[\ln(x+1)]^2}{2-y+4(x+1)[\ln(x+1)]^2}, \label{eq:32}\\
\sigma^{\prime} &=& 2\sigma\left[\frac{3-3x-3y+\sigma}{2-y+4(x+1)[\ln(x+1)]^2}-2\right]. \label{eq:33}
\end{eqnarray}
On a similar note, we can obtain the EoS parameters as,

\begin{eqnarray*}
\omega_{total} &=& -1+ \frac{6(x+y-1)-2\sigma}{3\left(2-y+4(x+1)[\ln(x+1)]^2\right)},\\
\omega_{DE} &=& \frac{-4(\sigma+3)[\ln(x+1)]^2-2 x\left(6 [\ln(x+1)]^2(x+1)+2\sigma[\ln(x+1)]^2+3\right)+y(\sigma-3)}{3(x+y)\left(2-y+4(x+1)[\ln(x+1)]^2\right)}.
\end{eqnarray*}
\begin{table}[H]
\renewcommand\arraystretch{1.5}
    \centering
    \begin{tabular}{|c|c|c|c|c|c|c|c|c|c|}
        \hline 
 ~~Name~~ & ~~Point/Curve~~ & ~~$\Omega_{m}~~$ & ~~$\Omega_{r}$~~ & ~~$\Omega_{DE}$~~ & ~~$q$~~ & ~~$\omega_{total}$~~ & ~~$\omega_{DE}$~~& ~~ Phase of Universe~~&~~Stability~~ \\
\hline \hline
$A_{2}$ & (0, 0, 1) & 0 & 1 & 0 & 1 & $\frac{1}{3}$ & - & Radiation dominated & Unstable Node\\
\hline
$B_{2}$ & (0, 0, 0) & 1 & 0 & 0 & $\frac{1}{2}$ & 0 & - & Matter dominated & ~~Unstable Saddle~~\\
\hline
$C_{2}$ & ($x$, $1-x$, 0) & 0 & 0 & 1 & -1 & -1 & -1 & ~~DE dominated~~ & Stable Node\\
\hline
    \end{tabular}
    \caption{Critical Points and the corresponding cosmology for {\bf Model-II}}
    \label{table:II}
\end{table}

\begin{figure}[H]
    \centering
   \includegraphics[scale=0.5]{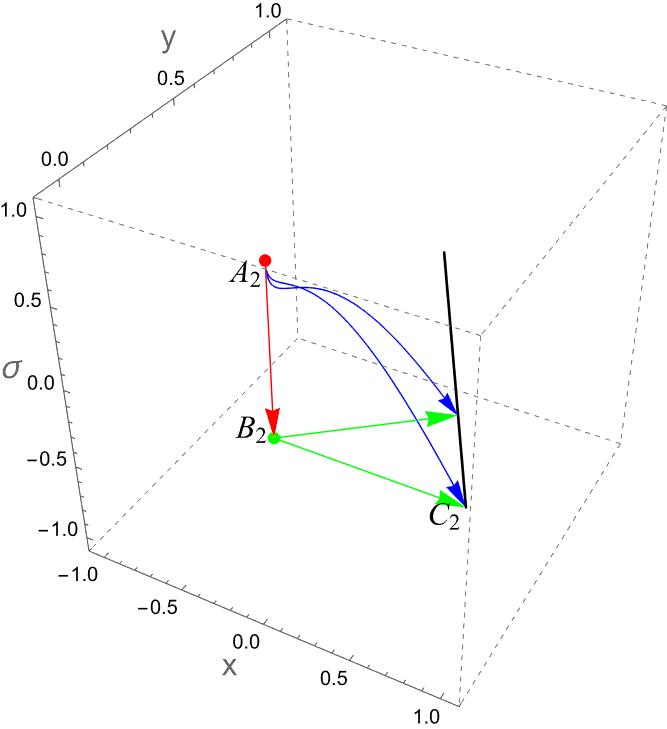}
    \includegraphics[scale=0.7]{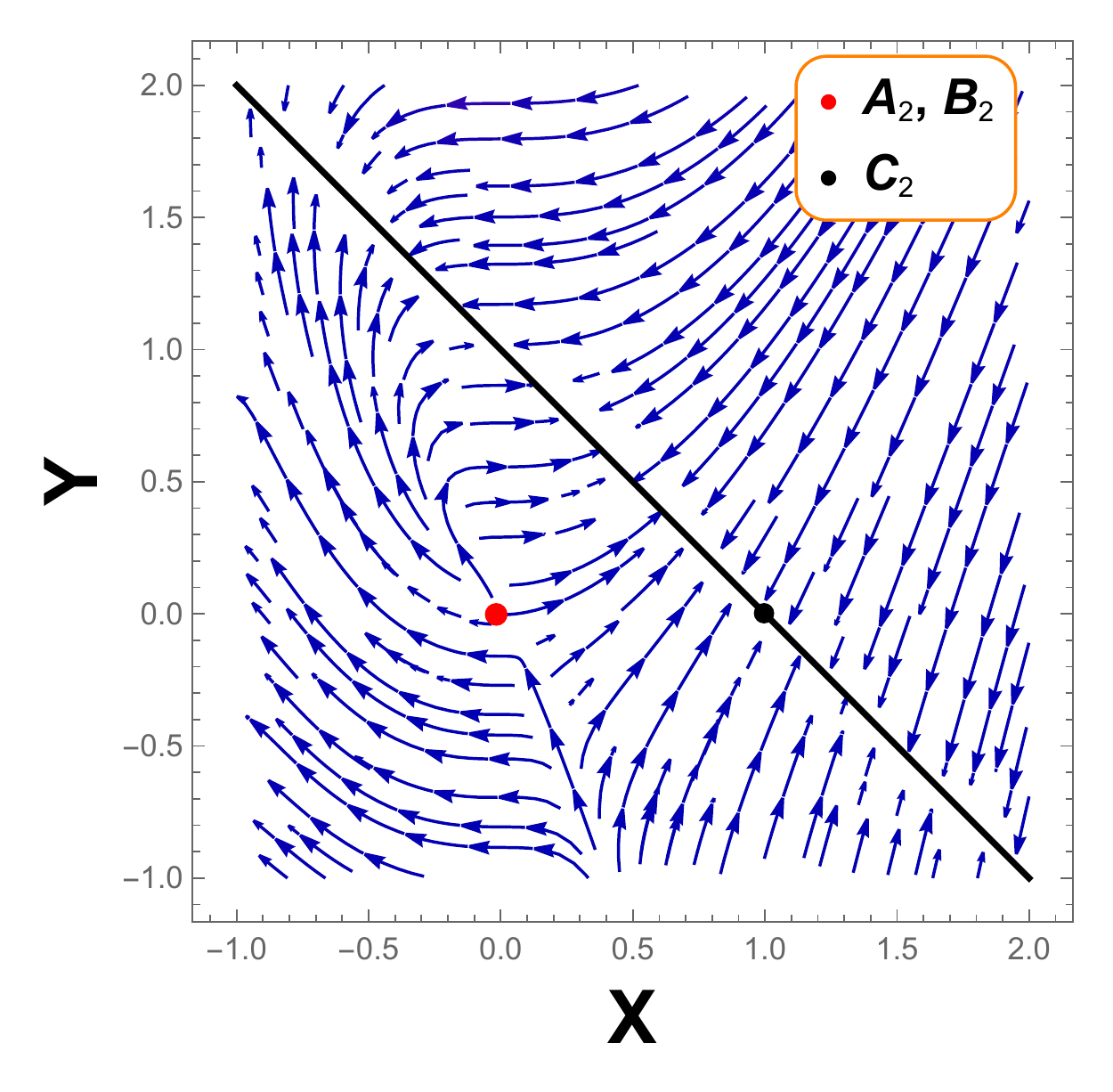}
    \caption{Phase-space trajectories on the $x$-$y$-$\sigma$ plane (Left Panel) and the portrait on the $x$-$y$ plane (Right Panel) for {\bf Model-II}.}
    \label{fig:VII}
\end{figure}

\begin{figure}[H]
\centering
\includegraphics[scale=0.7]{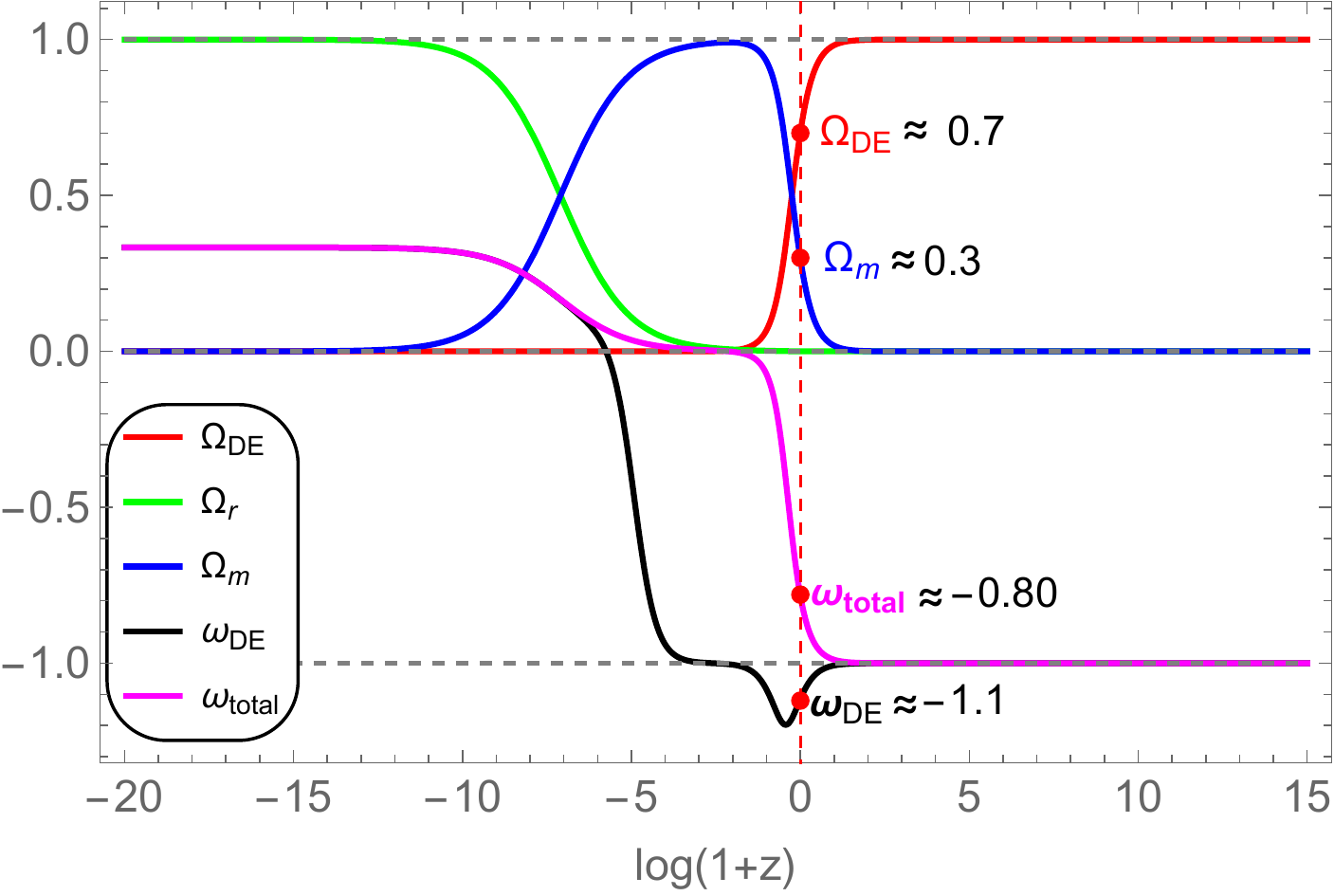}
\caption{Evolution of parameters for {\bf Model-II}. The initial conditions $x=10^{-15}$, $y=10^{-6}$ and $\sigma=10^{-1}$. The vertical dashed red line denotes the present time.}
\label{fig:VIII}
\end{figure}

TABLE \ref{table:II} provides the critical points and the cosmological behaviour at these points. The details description of each critical point has been narrated below. In FIG.- \ref{fig:VII}, the $2D$ and $3D$ phase portrait have been given to understand the stability of these points. 
\begin{itemize}
\item  \textbf{Critical Point $A_2$ (0, 0, 1) :} The critical point leads to the decelerating phase of the Universe, since the EoS parameter and deceleration parameter corresponding to this critical point is $\omega_{total}=\frac{1}{3}$ and $q=1$ respectively. The corresponding density parameter are $\Omega_{DE}=0$, $\Omega_{m}=0$ and $\Omega_{r}=1$. The eigenvalues for the corresponding critical point shows positive signature as given below. This critical point shows unstable behaviour. 
\begin{equation*}
\left\{4,~1,~0\right\}. 
\end{equation*}

\item \textbf{Critical Point $B_2$ (0, 0, 0) :} At this point, $\Omega_{DE}=0$, $\Omega_{m}=1$ and $\Omega_{r}=0$, i.e. the Universe shows matter dominated phase. The decelerated matter dominated Universe is confirmed by the corresponding value of the EoS parameter ($\omega_{total}=0$) and deceleration parameter $q=\frac{1}{2}$. Jacobian matrices with critical points have positive, negative real parts and zero eigenvalues. This critical point shows unstable saddle behaviour. The corresponding eigenvalues are given below:
    \begin{equation*}
    \{3,~-1,~~0\}.
    \end{equation*}

    \item \textbf{Curve of Critical Point $C_2$ (x, 1-x, 0) :} The corresponding EoS parameter is $\omega_{total}=-1$ and deceleration parameter is $q=-1$. This behaviour of the critical point leads to the accelerating phase of the Universe. Also, density parameters are $\Omega_{DE}=1$,  $\Omega_{m}=0$ and $\Omega_{r}=0$. This critical point is a stable node because it contains negative real part and zero eigenvalues of the Jacobian matrix.
    \begin{eqnarray*}
    \{0,~-4,~-3\}.
    \end{eqnarray*}
\end{itemize}
The phase portrait, which shows comparable trajectory plots, is an important tool in the study of dynamical systems. The stability of the models can be tested using the phase portrait. The phase portrait for system given in Eqns. \eqref{eq:28}-\eqref{eq:30} is shown in FIG.- \ref{fig:VII}. The (Left Panel) shows the trajectories in $x$-$y$-$\sigma$ ($3-D$) plane and (Right Panel) shows the trajectories in $x$-$y$ ($2-D$) plane, since $\sigma=0$ is an invariant sub-manifold. Here also the stability obtained along the curve $(x, 1-x, 0)$. This again shows the role of the dynamical variable $x$. Similar to {\bf Model-I}, the $3-D$ phase portrait for {\bf Model-II} shows attracting behaviour at critical curve $C_{2}$. All the repelling trajectories going away from critical points $A_{2}$ and $B_{2}$ and moving towards the stable curve $C_{2}$. From the evolution plot [FIG.- \ref{fig:VIII}], the present value of $\omega_{DE}=-1.1$ which is almost same as that of the value obtained through the scale factor.
\section{Results and Conclusion}\label{Sec.V}
To frame the cosmological models that concurs with the recent development of modern cosmology, several modified theories of gravity have been proposed. One among them is the symmetric teleparallel gravity or $f(Q)$ gravity. Two cosmological models are presented in this paper with some functional form of $f(Q)$ such as, (i) log-square-root model $\left[nQ_{0}\sqrt{\frac{Q}{\lambda_{1}Q_{0}}}\ln{\frac{\lambda_{1}Q_{0}}{Q}}\right]$ and (ii) exponential model $\left[Qe^{\frac{\mu\lambda_{2}}{Q}}-Q\right]$. In {\bf Model-I}, both $\frac{f(Q)}{H_{0}^{2}}$ and $f_{Q}(Q)$ decreases gradually whereas for {\bf Model-II}, $\frac{f(Q)}{H_{0}^{2}}$ increases and $f_{Q}(Q)$ decreases at late time. To analyse the dynamics of the Universe, we choose a time varying deceleration parameter that shows the early time deceleration and late time acceleration. {\bf Model-I} shows the quintessence behaviour at present time and behaves like the $\Lambda$CDM at late times whereas {\bf Model-II} shows the phantom behaviour and the present value of the DE EoS parameter obtained to be $\omega_{DE}=-0.82$ and $\omega_{DE}=-1.01$.\\

The dynamical system analysis has been performed for both the cosmological models obtained with the form of $f(Q)$ and the time varying deceleration parameter. For {\bf Model-I}, the critical points/curve are $A_{1}$ (0, 0, 1), $B_{1}$ (0, 0, 0) and $C_{1}$ ($x$, $1-x$, 0) given in Table-\ref{table:I}. The point $A_{1}$ is unstable node with all eigenvalues for Jacobian matrix positive real part and $B_{1}$ is unstable saddle with positive, negative real part and zero eigenvalues for Jacobian matrix. The curve $C_{1}$ has all eigenvalues for Jacobian matrix are negative real part and zero gives us the stable node behaviour. For more ideas about the critical point, the $2-D$ phase portrait has been shown in FIG.- \ref{fig:V} (Right Panel). FIG.- \ref{fig:VI} describes the evolutionary behavior of the Universe. At present $\Omega_{DE}\approx 0.7$ and $\Omega_{m}\approx 0.3$. Similarly for {\bf Model-II}, the critical points/curve are $A_{2}$ (0, 0, 1), $B_{2}$ (0, 0, 0) and $C_{2}$ ($x$, $1-x$, 0) given in Table-\ref{table:II}. Here also, $A_{2}$ and $B_{2}$ are unstable points, whereas $C_{2}$ shows the stable node behaviour. FIG.- \ref{fig:VIII} shows the present value of density parameters as, $\Omega_{DE}\approx 0.7$ and $\Omega_{m}\approx 0.3$. Both the models shows early radiation era and late time DE era, transition through matter dominated era. One can see from total EoS parameter obtained from hybrid scale factor and total EoS parameter from the evolution plot shows same behavior of Universe for both the models. \\

Regarding the unstable behaviour at the radiation and matter phase and stable behaviour obtained in the de Sitter phase, we wish to mention here that in some inflationary models of the early Universe, there can be instabilities associated with the rapid expansion of space during inflation. Whereas, the stability in the de-Sitter phase refers to the fact that it is associated with the positive cosmological constant or constant. The accelerated expansion of the Universe has been driven by the constant energy density. This leads to the stable stable de-Sitter phase, which is a desired feature for the accelerating behaviour. The results obtained in this paper are based on specific assumptions and model dependent and different behaviours can be obtained for different models and assumptions. Models that incorporate DE density or cosmological constant, in the phase portrait, the curve that passes through the stable points shows the trajectory of the de-Sitter phase. The models presented here exhibit similar behaviour.\\

Finally, we would like to mention here that the nonmetricity gravity can be used to obtain stable cosmological models and can able to describe the late time cosmic acceleration of the Universe. The value of EoS parameter has been important in addressing the evolutionary history of Universe. Scale factor and model parameters have been used in both models to constrain how well the dynamical parameters will behave. The stability of the models are analysed using the dynamical system and we get the stable points/curve for both the models.

\section*{Acknowledgement} BM acknowledges the support of IUCAA, Pune (India) through the visiting associateship program. The authors are thankful to the honorable referees for their valuable comments and suggestions for the improvement of the manuscript.

\end{document}